\documentstyle[multicol,prl,aps,epsf]{revtex}
\tightenlines
\begin{document}
\draft
\title{Direct Minimization Generating Electronic States with Proper Occupation Numbers}
\author{Kikuji Hirose and Tomoya Ono}
\address{Department of Precision Science and Technology, Osaka University, Suita, Osaka 565-0871, Japan}
\date{\today}
\maketitle
\begin{abstract}
We carry out the direct minimization of the energy functional proposed by Mauri, Galli and Car to derive the correct self-consistent ground state with fractional occupation numbers for a system degenerating at the Fermi level. As a consequence, this approach enables us to determine the electronic structure of metallic systems to a high degree of accuracy without the aid of level broadening of the Fermi-distribution function. The efficiency of the method is illustrated by calculating the ground-state energy of C$_2$ and Si$_2$ molecules and the W(110) surface to which a tungsten adatom is adsorbed.
\end{abstract}

\begin{multicols}{2}
\narrowtext

In the Kohn-Sham (KS) formulation of density-functional theory \onlinecite{hks,janak}, the energy functional of a system of electrons is stationary at the point of the ground-state energy with respect to the variation of an arbitrary set of single-particle orthonormal wave functions \{$\psi_i$\} and occupation numbers \{$n_i$\} where $n_i$ varies between zero and one. The Janak theorem \cite{janak} implies that the occupation numbers of the ground state at zero temperature are chosen such that $n_i=1$ for the states of KS eigenvalues below the Fermi level, and $n_i=0$ for those above the Fermi level. For systems with degeneracies at the Fermi level such as metallic systems, however, the determination of the occupation numbers of the degenerate states is not trivial \onlinecite{brute,pederson}. One popular approximation has been to determine the occupation numbers of the states near the Fermi level using the Fermi-distribution (FD) or its substitute function where a fictitious temperature, a parameter to adjust the broadening of levels, is included \onlinecite{fddf1,fddf2}. Further, the thermodynamic free energy is minimized instead of the electronic total energy. This procedure has the advantage that the calculations of electronic wave functions for metallic systems typically converge more stably and rapidly, and that the broadening of each level roughly imitates larger systems or a better sampling in the Brillouin zone. In principle, however, no physical significance is given to temperature or entropy in the context of the zero-temperature KS scheme. It is even more serious in practice that theoretical results depend on the chosen values of the fictitious broadening temperature, and that all of the occupation numbers of the degenerate states at the Fermi level are restricted to be numerically equal, which is usually incorrect for accidental degeneracies.

In this Letter, we demonstrate that the direct minimization (DM) of the energy functional proposed by Mauri, Galli and Car (MGC) \cite{galli} leads to the assignment of the proper occupation numbers to both degenerate and nondegenerate states; moreover, the DM procedure yields the correct self-consistent solutions of the KS equation without usage of conventional self-consistent field techniques. Consequently, the electronic structure of metallic systems at zero temperature can be obtained to a high degree of accuracy with no aid of a FD broadening. We combine the DM method with the real-space finite-difference method utilizing the timesaving double-grid technique\cite{tsdg}. The real-space calculation methods have tackled serious drawbacks of the plane-wave approach, e.g., its inability to describe strictly nonperiodic systems such as clusters and solid surfaces.

The MGC energy functional for an $N$-electron system is written as
\begin{eqnarray}
\label{eqn:energyfunctional1}
E[\{\phi\},\eta] &=& \sum_{\sigma=\uparrow \downarrow} \sum_{i,j}^M Q^\sigma_{ij} \langle \phi^\sigma_j| -\frac{1}{2}\nabla^2 | \phi^\sigma_i \rangle \nonumber \\
&& + F[\rho^\uparrow,\rho^\downarrow] + \eta\left\{ N - \int \rho({\bf r}) d {\bf r} \right\} ,
\end{eqnarray}
where $\{ \phi_i^\sigma \}$ is an arbitrary set of $M$ linearly independent {\it overlapping} wave functions with spin index $\sigma$, which are assumed here to be real functions for simplicity, and $M$ is taken to be not smaller than the number of the occupied states for each spin. $F[\rho^\uparrow,\rho^\downarrow]$ is the sum of the external, Hartree, and exchange-correlation potential energy functionals, $\eta$ is the electronic chemical potential, $Q^\sigma$ is an ($M \times M$) matrix: $Q^\sigma_{ij}=2\delta_{ij}-S^\sigma_{ij}$, and $S^\sigma$ is the overlap matrix: $S_{ij}^\sigma=\langle \phi_i^\sigma|\phi_j^\sigma \rangle$. The charge density is defined as\\
\begin{equation}
\label{eqn:density1}
\rho({\bf r})=\sum_{\sigma=\uparrow \downarrow} \rho^\sigma({\bf r}) \: \: \mbox{and} \: \: \rho^\sigma({\bf r})=\sum_{i,j}^M Q^\sigma_{ij} \phi_j^\sigma({\bf r})\phi_i^\sigma({\bf r}) .
\end{equation}
The form of the energy functional Eq. (\ref{eqn:energyfunctional1}) was originally introduced for computation with linear system-size scaling $O(N)$, and each wave function $\phi_i^\sigma$ was approximated to be a Wannier-like function localized in an appropriate region of space (i.e., a localized orbital) to reduce the amount of computation. In this paper, we adhere to ordinary {\it extended} wave functions, being free from the errors caused by the localization of wave functions.

$E[\{\phi\},\eta]$ is minimized by a steepest-descent or conjugate-gradient algorithm without constraint of the orthogonalization of wave functions. The derivative of the functional with respect to the function $\phi_i^\sigma$ is required for the minimization, which is given by
\begin{eqnarray}
\label{eqn:def1}
\frac{\delta E[\{\phi\},\eta]}{\delta \phi_i^\sigma} &=& 2 \sum_{j}^M \left[ ( \hat{H}_{KS}^\sigma[\{ \phi\}] - \eta ) | \phi_j^\sigma \rangle Q_{ji}^\sigma \right. \nonumber \\
& & \left. - |\phi_j^\sigma \rangle \langle \phi_j^\sigma| (\hat{H}^\sigma_{KS}[\{\phi\}] - \eta ) |\phi_i^\sigma \rangle \right] ,
\end{eqnarray}
where $\hat{H}^\sigma_{KS}[\{\phi\}]$ is the KS Hamiltonian. It is straightforward to verify that the ground-state energy is a stationary point of $E[\{\phi\},\eta]$. Consider the following set of the ground-state wave functions \{$\phi_i^{\sigma 0}$\}: $|\phi_i^{\sigma 0}\rangle=a_i^\sigma|\chi_i^\sigma\rangle$, where $|\chi_i^\sigma\rangle$ is the normalized eigenfunction of $\hat{H}_{KS}^\sigma$ at $\{\phi^\sigma_i\}=\{\phi^{\sigma 0}_i\}$ with the eigenvalue $\varepsilon_i^\sigma$, i.e., $\hat{H}^\sigma_{KS}[\{\phi^{\sigma 0}_i \}]|\chi_i^\sigma\rangle=\varepsilon_i^\sigma|\chi_i^\sigma\rangle$, and the coefficient $a_i^\sigma$ is set such that $|a_i^\sigma|=1$ for $\varepsilon_i^\sigma<\eta$, $0\leq|a_i^\sigma|\leq 1$ for $\varepsilon_i^\sigma=\eta$, and $a_i^\sigma=0$ for $\varepsilon_i^\sigma>\eta$. Obviously, $\{\phi_i^{\sigma 0}\}$ is a stationary point of $E[\{\phi\},\eta]$, since
\begin{eqnarray}
\delta E[\{\phi\},\eta]/\delta \phi_i^\sigma|_{\{\phi_i^\sigma\}=\{\phi_i^{\sigma 0}\}}&=&4a_i^\sigma(1-|a_i^\sigma|^2)(\varepsilon_i^\sigma-\eta)|\chi_i^\sigma \rangle \nonumber \\
&=&0.
\end{eqnarray}

We now show that the MGC energy functional Eq. (\ref{eqn:energyfunctional1}) is identical to the functional in the standard form
\begin{eqnarray}
\label{eqn:energyfunctional2}
E[\{\psi\},\eta] &=& \sum_{\sigma=\uparrow \downarrow} \sum_i^M n^\sigma_i \langle \psi^\sigma_i | -\frac{1}{2}\nabla^2 | \psi^\sigma_i \rangle \nonumber \\
&& + F[\rho^\uparrow,\rho^\downarrow] + \eta\left\{ N - \int \rho({\bf r}) d {\bf r} \right\} ,
\end{eqnarray}
with
\begin{equation}
\label{eqn:density2}
\rho({\bf r}) = \sum_{\sigma=\uparrow \downarrow} \rho^\sigma({\bf r}) \: \: \mbox{and} \: \: \rho^\sigma({\bf r})= \sum_i^M n_i^\sigma |\psi_i^\sigma ({\bf r})|^2 ,
\end{equation}
where $\{\psi_i^\sigma\}$ and \{$n_i^\sigma$\} are sets of {\it orthonormal} wave functions and the corresponding occupation numbers, respectively, but $n^\sigma_i$ varies in the range that $-\infty < n^\sigma_i \leq 1$. The proof is as follows: A set of \{$\phi_i^\sigma$\} in Eqs. (\ref{eqn:energyfunctional1}) and (\ref{eqn:density1}) is transformed to an orthonormal set $\{\tilde{\psi}_i^\sigma\}$ using an orthogonalization algorithm, e.g., the Gram-Schmidt method, and then $\phi_i^\sigma$ is represented as $\phi_i^\sigma=\sum_l^M c_{l,i}^\sigma \tilde{\psi}_l^\sigma$. Substituting this expansion into Eqs. (\ref{eqn:energyfunctional1}) and (\ref{eqn:density1}), we obtain for the energy functional
\begin{eqnarray}
\label{eqn:energyfunctional3}
E[\{\tilde{\psi}\},\eta] &=& \sum_{\sigma=\uparrow \downarrow} \sum_{k,l}^M P^\sigma_{kl} \langle \tilde{\psi}^\sigma_l| -\frac{1}{2}\nabla^2 | \tilde{\psi}^\sigma_k \rangle \nonumber \\
&& + F[\rho^\uparrow,\rho^\downarrow]+ \eta\left\{ N - \int \rho({\bf r}) d {\bf r} \right\} ,
\end{eqnarray}
and a similar expression for the charge density, where
\begin{equation}
P_{k l}^\sigma = \sum_{i,j}^M Q_{ij}^\sigma c_{k,i}^\sigma c_{l,j}^\sigma = 2 T_{k l}^\sigma - \sum_{m}^M  T_{k m}^\sigma T_{m l}^\sigma,
\end{equation}
with $T_{k l}^\sigma=\sum_{i}^M c_{k,i}^\sigma c_{l,i}^\sigma$. Here, the matrix $T^\sigma$, of which the element of the $k$ row and the $l$ column is $T_{k l}^\sigma$, is a non-negative definite Hermitian matrix, which can be diagonalized by a unitary matrix $U^\sigma$ as $(U^{\sigma \dagger} T^\sigma U^\sigma)_{k l}=\delta_{k l} \lambda_k^\sigma$ with $\lambda_k^\sigma$ being a non-negative eigenvalue of $T^\sigma$. Hence,
\begin{equation}
\label{eqn:density4}
(U^{\sigma \dagger} P^\sigma U^\sigma)_{k l} = \delta_{k l}(2-\lambda_k^\sigma)\lambda_k^\sigma.
\end{equation}
Defining $\psi_i^\sigma$ and $n_i^\sigma$ as $\psi_i^\sigma=\sum_j^M \tilde{\psi}_j^\sigma U_{ji}^\sigma$ and $n_i^\sigma = (2 - \lambda_i^\sigma) \lambda_i^\sigma$, we obtain Eq. (\ref{eqn:energyfunctional2}) from Eq. (\ref{eqn:energyfunctional3}), and the inequality $-\infty < n^\sigma_i \leq 1$. (Q.E.D.)

\begin{figure}
\epsfbox{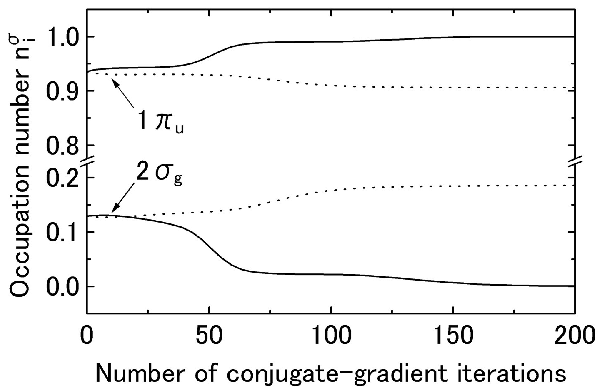}
\caption{Occupation number $n_i^\sigma\equiv(2-\lambda_i^\sigma)\lambda_i^\sigma$ as a function of the number of conjugate-gradient iterations. (For the definition of $\lambda_i^\sigma$, see text preceding Eq. (\ref{eqn:density4}).) The solid (dotted) curves represent the occupation numbers in the case of the atomic separation $R=2.30$ a.u. ($2.50$ a.u.); each run starts from the electronic structure at $R=2.40$ a.u.}
\label{fig:2}
\end{figure}

It is very important that the occupation numbers defined above do not exceed one, because this fact means that in the course of the minimization of the the MGC energy functional Eq. (\ref{eqn:energyfunctional1}), the Pauli principle automatically works to prevent more than one electron from falling into each single-particle level. The unphysical negative occupation gives rise to no severe problem in practical calculations, as long as initial wave functions for iterations are chosen to be close to the ground-state solution. In this case, the occupation numbers maintain non-negative values during iterations, which is assured by experience (e.g., see Fig. ~\ref{fig:2}).

The overall computational scaling in our DM procedure combined with the real-space finite-difference method of Ref. \cite{tsdg} amounts to $O$($M^2N_{grid}$) operations in the calculations of the energy functional Eq. (\ref{eqn:energyfunctional1}) and its derivative Eq. (\ref{eqn:def1}), since the $\phi^\sigma_i$'s are now assumed to be {\it extended} wave functions. Here, $N_{grid}$ is the number of coarse grid points in real space. It is noted that $O$($M^2N_{grid}$) is equal to the scaling order in the orthogonalization of wave functions indispensable in the conventional approach of solving the KS equation by an iterative algorithm, and that the dominant scaling in the calculations of the energy functional and its derivative based on a plane-wave basis set is also $O$($M^2N_{basis}$) with $N_{basis}$ being the number of plane-wave basis functions, when advantage is taken of the fast Fourier transform technique.

In order to illustrate the utility of the DM method and the difficulty in the usage of the FD function, the electronic structures of C$_2$, Si$_2$ and a tungsten adatom  adsorbed on the W(110) surface are calculated. Hereafter, we obey the nine-point finite-difference formula\cite{chelikowsky} for the derivative arising from the kinetic-energy operator, and the dense-grid spacing is fixed at $h_{dens}=h/3$ \cite{tsdg}, where $h$ denotes the coarse-grid spacing. The norm-conserving pseudopotential is employed in a Kleinman-Bylander nonlocal form \onlinecite{pseudop,kobayashi}.

\begin{figure}
\epsfbox{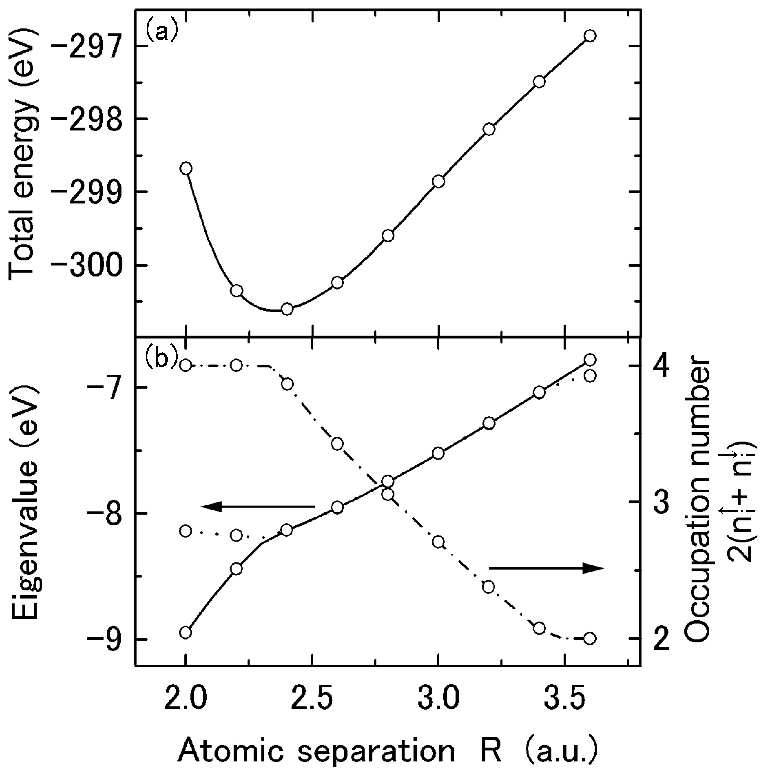}
\caption{C$_2$ adiabatic-potential curve, eigenvalues of $1\pi_u$ and $2\sigma_g$ states, $1\pi_u$ occupancy as a function of atomic separation $R$. In (b), the solid (dotted) curve represents the eigenvalue of $1\pi_u$ ($2\sigma_g$) state, and the dash-dotted curve shows the occupation numebr of $1\pi_u$ state. Circles correspond to data obtained by the brute-force method.}
\label{fig:1}
\end{figure}

We show in Figs. ~\ref{fig:1} and ~\ref{fig:2} the results of the application to the carbon dimer as a demonstration of the advantages of our method. It is well-known that a self-consistent screening potential for the carbon dimer cannot be stabilized by the integral occupation of the occupied states. Pederson and Jackson \cite{pederson} showed by the study based on the local-density approximation \cite{lda} that the minimum energy corresponds to a fractionally occupied system in the range of the atomic separation $R$=2.4 - 3.7 a.u. and that KS eigenvalues for the fractionally occupied states are degenerate. In Figs. ~\ref{fig:1} and ~\ref{fig:2}, we took the grid spacing $h$=0.33 a.u. in a cell of 16$^3$ a.u. under the nonperiodic boundary condition (b.c.) of vanishing wave function, and set the number of electrons $N$=8 and the number of wave functions $M$=5. Exchange-correlation effects were treated with the local-density approximation according to Pederson and Jackson \cite{pederson}. Figure~\ref{fig:1} illustrates the ground-state total energy, the KS eigenvalues of the $1\pi_u$ and $2\sigma_g$ states, and the occupation number of the $1\pi_u$ state as a function of atomic separation. The result of the brute-force method \cite{brute} implementing the direct variation of the occupation numbers is also depicted here for comparison with our DM method. The total energies predicted by both methods coincide with one another [see Fig. ~\ref{fig:1}(a)]. As shown in Fig.~\ref{fig:1}(b), KS eigenvalues for the $1\pi_u$ and $2\sigma_g$ states are degenerate in the range $R$=2.34 - 3.45 a.u. and the occupation numbers for these states become fractional in this range, which accords with the results of the other methods \onlinecite{brute,pederson}. In Fig.~\ref{fig:2}, we plot the occupation numbers of these states at $R=2.30$ a.u. and $2.50$ a.u. as a function of the number of conjugate-gradient iterations, where the electronic structure at $R=2.40$ a.u. is used as the starting point. One can see that the iterative process is markedly stable and our DM procedure gives good convergence of the occupation numbers.

\begin{figure}
\epsfbox{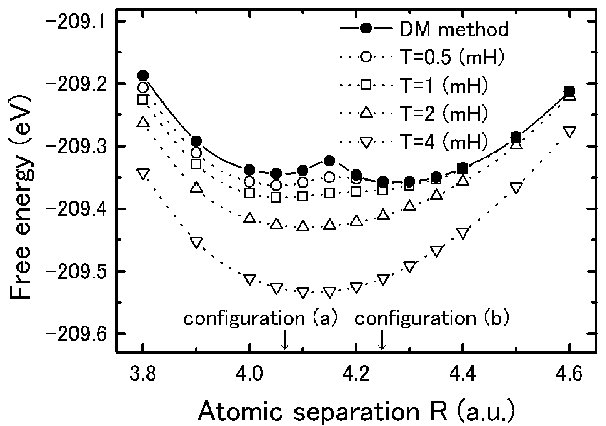}
\caption{Si$_2$ adiabatic-potential curves as a function of atomic separation $R$ obtained by the DM method and the conventional KS scheme at different broadening temperatures T. The indicated configurations are (a) $(1\sigma_g)^2(1\sigma_u)^2(1\pi_u)^4$ and (b) $(1\sigma_g)^2(1\sigma_u)^2(2\sigma_g)^2(1\pi_u)^2$.}
\label{fig:3}
\end{figure}

We next give an example that the FD method including a fictitious broadening temperature leads to an incorrect ground-state geometry in molecular-dynamics simulations. Figure ~\ref{fig:3} shows the curves of the thermodynamic free energy versus the atomic separation for the silicon dimer. The calculation was performed under the following conditions: the grid resolution $h=0.50$ a.u., a cell of 24$^3$ a.u. under the nonperiodic b.c. and the local-spin-density approximation \cite{lda} for the exchange-correlation potential. The experiments \cite{exp} proved that the energy difference between two molecular configurations (a) and (b) indicated in Fig. ~\ref{fig:3} is quite small, the equilibrium bond length is 4.07 a.u. for the configuration (a) and 4.25 a.u. for (b), and the ground state with the lowest total energy is the latter. Some theoretical analyses have been carried out to examine the situation \onlinecite{brute,chelikowsky,yin}. As seen in Fig.~\ref{fig:3}, our DM procedure can yield results that are in excellent agreement with the empirical data. On the contrary, the conventional KS scheme using the FD function is unable to search out the minimum configuration (b) in molecular-dynamics iterations at the broadening temperature T$\geq$1 mH.

\begin{figure}
\epsfbox{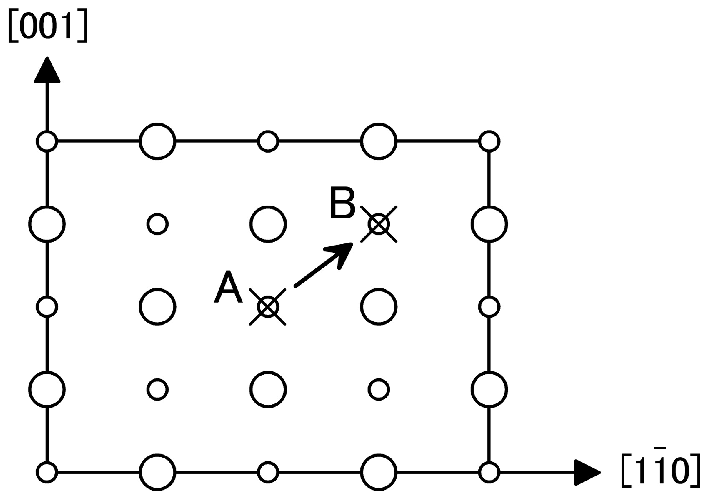}
\caption{Top view of W(110) surface-layer atoms (large open circles), second-layer atoms (small open circles) and adatom (crosses) for the surface diffusion discussed in the text.}
\label{fig:4}
\end{figure}

As a final example, we evaluate the energy barrier for a tungsten adatom to hop along the [1$\bar{1}$1] direction on the W(110) surface. Our test system is a cell of 17 tungsten atoms with 102 electrons, consisting of one adatom and two rigid W(110) planes, under the periodic b.c. at the [001] and [1$\bar{1}$0] directions, and the nonperiodic b.c. of vanishing wave function at the [110] one (i.e., thin film model). The number of wave functions $M=64$ for each spin, the coarse-grid spacing $h=0.30$ a.u., and the local-spin-density approximation were employed. Only the $\Gamma$ point of the Brillouin zone was sampled. We first evaluate the ground-state geometry for the adatom located at point A in Fig.~\ref{fig:4} using our DM approach, and then displace the adatom along the [1$\bar{1}$1] direction from point A to B. The diffusion barrier of the tungsten adatom is given in Table \ref{tbl:1}. The numerical error in the barrier at T=2 mH is found to be not negligible but about 10 \% of the true value obtained from our DM scheme.

\narrowtext
\begin{table}
\caption{The diffusion barrier of a tungsten adatom on the rigid W(110) surface along the [1$\bar{1}$1] direction. Data from our DM method and the KS scheme including a broadening temperature are presented.}
\begin{tabular}{cc}
Temperature (Hartree)&Diffusion barrier (eV) \\ \hline
Present work & 1.40 \\
0.001 & 1.43 \\
0.002 & 1.54 \\
0.004 & 1.54 \\
\end{tabular}
\label{tbl:1}
\end{table}

In conclusion, we have demonstrated a new variational technique to introduce appropriately the fractional occupation numbers, and calculated the ground-state energy and the charge density to a high degree of accuracy for a system with degeneracies at the Fermi level. Our scheme requires no additional statistical parameter such as a broadening temperature. Moreover, when a system is assumed to be described in terms of localized wave functions, our procedure makes it possible to perform the calculation with linear system-size scaling $O(N)$. Research in this direction is in progress.

One of the authors (T.O.) is a Research Fellow of the Japan Society for the Promotion of Science. This work was partially supported by a Grant-in-Aid for COE Research. The numerical calculation was carried out by the computer facilities at the Institute for Solid State Physics at the University of Tokyo.

\end{multicols}
\end{document}